\documentclass[
 preprint,
 amsmath,amssymb,
 aps,
 pra,
]{revtex4-1}

\usepackage{units}
\usepackage{upgreek}
\usepackage{graphicx}
\usepackage{dcolumn}
\usepackage{bm}
\usepackage{dsfont}

\begin{document}

\title{Shaping caustics into propagation-invariant light}

\author{Alessandro Zannotti}
 \email{a.zannotti@uni-muenster.de}
\author{Cornelia Denz}
\affiliation{
 Institute of Applied Physics and Center for Nonlinear Science (CeNoS), University of Muenster, 48149 Muenster, Germany
}
\author{Miguel A. Alonso}
\affiliation{
Aix Marseille Univ, CNRS, Centrale Marseille, Institut Fresnel, UMR 7249, 13013 Marseille, France,\\
The Institute of Optics, University of Rochester, Rochester NY 14627, USA
}
\author{Mark R. Dennis}
\affiliation{
 School of Physics and Astronomy, University of Birmingham, Birmingham B15 2TT, UK \\
 H H Wills Physics Laboratory, University of Bristol, Bristol BS8 1TL, UK
}

\date{\today}

\maketitle

\section*{Abstract}

Structured light has revolutionized optical particle manipulation and nano-scale material processing. In particular, propagation-invariant structured light fields, such as Bessel beams, have enabled applications that require robust intensity distributions. Their self-healing nature facilitates imaging with enhanced resolution e.g.~in light-sheet microscopy.
The prominent high-intensity features of propagation-invariant fields such as Airy, Bessel, and Mathieu beams can be understood in terms of caustics. While these beams have found many applications in material processing and trapping, these technologies would greatly benefit from structured, controllable intensities in a variety of shapes well beyond the standard families of propagation-invariant beams. 
Here we generalize propagation-invariant beams by tailoring their caustics through two different methods. We illustrate these approaches by implementing various tailored propagation-invariant beams experimentally, whose patterns range from simple geometric shapes to complex configurations such as words.
This approach clarifies that the known solutions are a small subset of a far more general set of propagation-invariant fields with intensity maxima concentrated around any desired curve.

\newpage
\cleardoublepage

\section*{Introduction}
The field of structured light has grown significantly since the early studies of Laguerre- and Hermite-Gaussian modes in laser cavities~\cite{Andrews2011,Rubinstein-Dunlop2017}. This growth stemmed both from increased theoretical understanding and from the advent of new optical devices such as spatial light modulators (SLMs).
This area of research has now transcended optics, leading to a range of applications in fundamental physics~\cite{Hansen2016}, telecommunications~\cite{Wang2012, Torres2012, Xie2018}, security~\cite{Sit2017,Sit2018,Erhard2018}, micro-machining~\cite{Mathis2012, Courvoisier2016}, imaging~\cite{Hell1994,Willig2007,Fahrbach2010,Vettenburg2014}, and the manipulation of cells and microorganisms~\cite{Dholakia2011, Woerdemann2013}. 
In particular, the fact that phase-structured light carries orbital angular momentum (OAM)~\cite{Allen1999} allows the exertion not only of forces but also of torques onto atoms, bacteria, or micro-machines~\cite{Andrews2011, Rubinstein-Dunlop2017}. 

A class of structured optical field that has received considerable attention is that of the so-called \emph{propagation-invariant} or \emph{non-diffracting} beams, whose transverse intensity distribution remains essentially invariant over a significant propagation distance~\cite{Bouchal2003a}. Propagation-invariant beams have a transverse angular spectrum confined to a ring~\cite{Whittaker1903} whose radius is proportional to their numerical aperture. 
The best-known examples are Bessel~\cite{Durnin1987a,Durnin1987} , Mathieu~\cite{Bouchal2003a, Rose2012, Acias2018}, and Weber~\cite{Bouchal2003a,Rose2012,Sosa-Sanchez2017} beams, which are described by closed-form expressions that are separable in polar, elliptic and parabolic coordinates, respectively. The intensity maxima of these beams are therefore localized around the corresponding conic section shapes, characterized by their caustics~\cite{Berry1979, Siviloglou2007, Alonso2017, Acias2018}. 
Some of these propagation-invariant beams have been used in advanced optical trapping setups~\cite{Garces-Chavez2002, Dholakia2011, Woerdemann2013}, imaging with high resolution~\cite{Fahrbach2010, Vettenburg2014}, and ultrafast nano-scale material processing with high aspect ratios~\cite{Mathis2012, Courvoisier2016, Meyer2017}. 
However, having only a limited set of beams of this type restricts significantly their usefulness. All the applications mentioned earlier would greatly benefit from the ability to tailor the shapes of propagation-invariant beams for the purpose at hand.

The goal of this work is to present two methods for the design of propagation-invariant beams whose main transverse intensity features trace any desired curved shape by shaping corresponding caustics into light fields. These methods are illustrated with the experimental implementation of beams whose intensity features trace a range of geometrical shapes as well as more complex patterns such as words, which remain essentially invariant over a significant propagation distance.

\section*{Results}
\subsection*{Caustics in propagation-invariant beams}
The conceptual basis for this work is the relation between these wave solutions and the simpler ray model, for which the intensity maxima follow the shapes of the {\it caustics}, which are the envelopes of the two-parameter family of rays associated with the field~\cite{Berry1980}. Within the ray picture, propagation invariance requires that all rays travel at the same angle $\theta$ with respect to the propagation direction (chosen as the $z$ axis), and that the two-parameter ray family is composed of one-parameter subfamilies of parallel rays constrained to planes parallel to $z$. This structure guarantees that the caustics themselves are invariant in $z$, since they correspond to the envelopes of these planes. The fact that all rays have equal angle with respect to the $z$ axis implies that the transverse Fourier spectrum of the wave field is restricted to a ring~\cite{Bouchal2003a} (whose radius is $k_\bot=k\sin\theta$, where $k$ is the wavenumber). The azimuthal angle $\phi$ of each sub-family of rays indicates the point along the ring to which they correspond.

The mathematical correspondence between the ring's amplitude and phase distributions, $A(\phi)$ and $\Phi(\phi)$, and the beam's transverse field is given by Whittaker's integral~\cite{Whittaker1903,Bouchal2003a}:
\begin{equation}
\psi(\mathbf{r}) = \oint A(\phi) \exp\left[\text{i}\Phi(\phi) + \text{i}k_\bot\mathbf{r}\cdot\mathbf{u}(\phi)\right] \text{d}\phi,
\label{eq:Whittaker}
\end{equation}
where $\mathbf{r}=(x,y)$ is the transverse position vector and $\mathbf{u}(\phi) = (\cos\phi, \sin\phi)$ is a unit vector that indicates the transverse direction of the corresponding rays and the planes that embed them. 
The connection with ray-optics and caustics results from applying the method of stationary phase~\cite{Berry1980} to Eq.~(\ref{eq:Whittaker}). As shown in the Methods section, this leads to the following relation in terms of the parametrized caustic shape~$\mathbf{r}_{\rm c}(\phi)$: 
\begin{align}
   \mathbf{r}_{\rm c}(\phi) &= \frac{1}{k_\bot}\left[\Phi(\phi)''\mathbf{u}(\phi) - \Phi'(\phi)\mathbf{u}'(\phi)\right].
\label{eq:rc}
\end{align}
By choosing $A$ and $\Phi$ appropriately, different caustic shapes can be achieved.
Turning points --- local maxima and minima --- in the continuous modulation of this phase function $\Phi$ correspond to cusp caustics. The beam's intensity maxima are localized within the vicinity of these caustics. 
For example, Bessel beams (for which $A$ is constant and $\Phi(\phi)=\ell\phi$ for integer $\ell$) have caustics with circular (or punctual, for $\ell=0$) cross-sections~\cite{Andreev1991,Palchikova1998,Bliokh2010a,Alonso2017}, while Mathieu beams have caustics with elliptic or hyperbolic cross-sections~\cite{Acias2018}.

\begin{figure}[t]
\centering
\includegraphics[width=1\columnwidth]{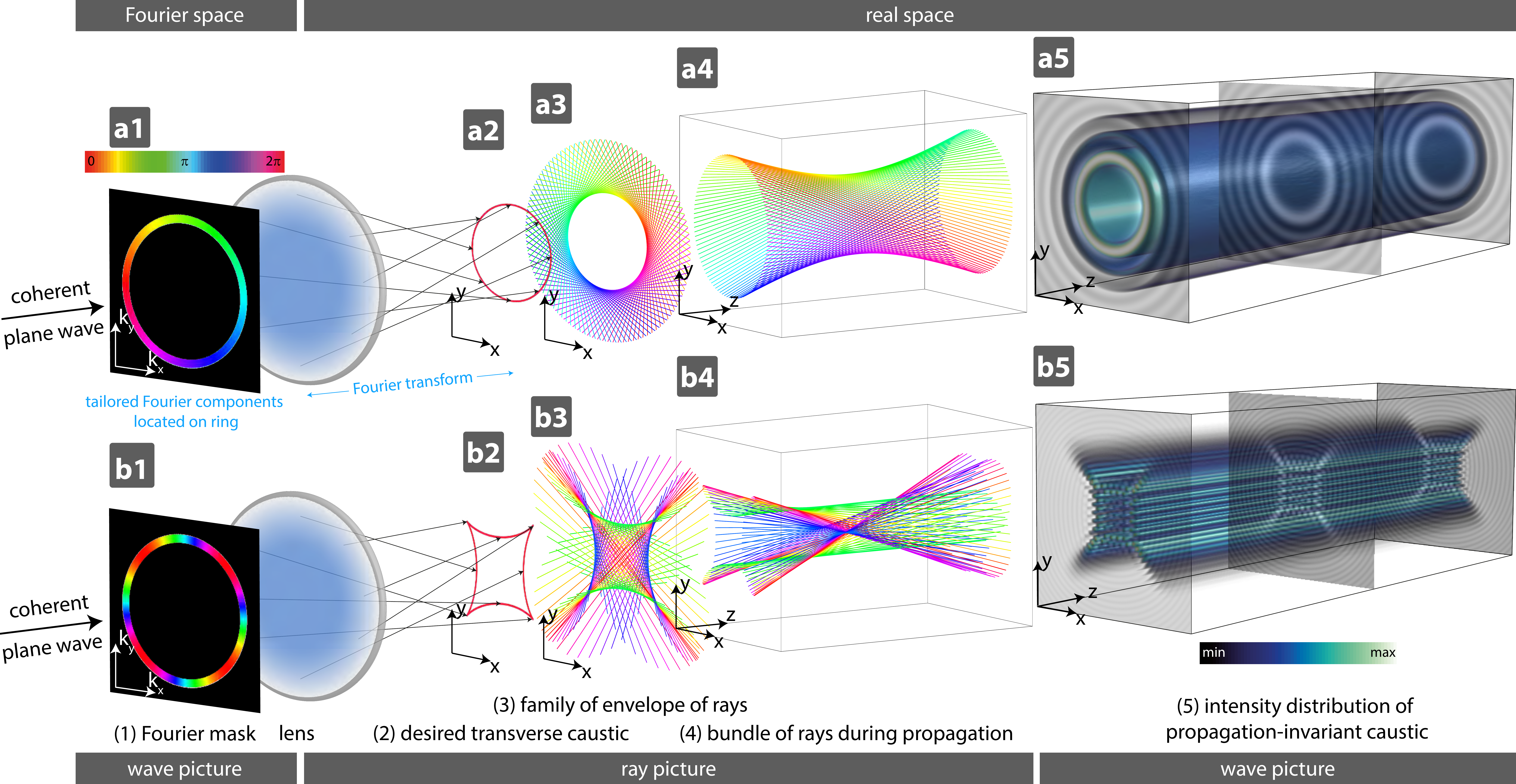}
\caption{
	Caustics in propagation-invariant beams.
	(a) A Bessel beam $J_1$ with OAM of charge $\ell = 1$ has a circular caustic. (a1) Fourier phase pattern, with azimuthally linearly increasing phase $\Phi = \ell\phi$, confined to an infinitesimal thin ring. Fourier transform by a lens. Fourier wavefront shaping determines the direction of the propagating rays, and the caustic (a2) forms in real space as the envelope of the rays in a transverse plane (a3). A subset of the rays, each coming from a specific point on the ring, is shown in (a4). The complete family of rays is composed of a continuum of bundles like the one shown in (a4) but displaced in $z$. The corresponding field has a propagation-invariant transverse intensity (a5) with a pronounced central ring according to the circular caustic. (b) A more complex phase function $\Phi = \ell\phi -q/2\sin(2\phi)$ yields an astroid caustic, which includes four cusps related to the four turning points of the phase function.}
\label{fig:Idea}%
\end{figure}


Figure~1a illustrates some of these ideas for the simple example of a Bessel beam with topological charge $\ell = 1$. 
The transition between the Fourier plane and the physical space is implemented experimentally with a converging lens placed one focal length away from the Fourier plane. 
The cross-sections of the planes containing the rays~(a3) form a circular caustic~(a2). The 3D configuration of some of these rays (one from each subfamily) is also shown~(a4); the complete two-parameter family of rays includes the subset shown as well as replicas displaced in $z$ (see the Supplementary Materials for a discussion of the whole ray family).
Note that the rays are color-coded (by hue) to show the corresponding phase at the ring. For visual convenience the accumulation of phase due to propagation is factored out. 
Finally, the 3D intensity distribution for this Bessel beam, dominated by the innermost intensity circle, is also shown~(a5).
Figure~1b shows similar information for a more complicated beam with $\Phi(\phi) = -2.5\sin(2\phi)$. This phase distribution leads to a propagation-invariant astroid caustic that exhibits four cusps, which correspond to the four turning points of the phase function over the Fourier ring. 
More information regarding the visualization of the rays is provided in the Supplementary Material.

\subsection*{Solving the inverse problem using the differential equation \ref{eq:rc}}

Our goal, however, is to solve the inverse problem: find the phase distribution $\Phi$ that produces any desired propagation-invariant caustic. 
The first approach is the solution of Eq.~(\ref{eq:rc}) for $\Phi(\phi)$ for a given $\mathbf{r}_{\rm c}(\phi)$. 
For simplicity we keep the amplitude $A(\phi) $ constant. The details of this approach are given in the Methods section, and some results are shown in Fig.~2 for simple caustic configurations.
For example, solving the differential equation for the astroid caustic (a1) gives the sinusoidal phase function $\Phi(\phi) = \ell\phi - q/2\,\sin(2\phi)$ (a2), with $q \in \mathds{R}$, which is mapped mod $2\pi$ on the Fourier ring (a3). (Note that this astroid can carry OAM with topological charge~$\ell$.)
From the Fourier phase function, we calculate the transverse ray picture of the beam (a4). 
The resulting experimentally-measured transverse intensity (a5) and phase (a6) distributions (for $q=5$) show a square lattice at the centre, surrounded by several rings. 
This beam interpolates between Bessel beams ($q=0$) and four interfering plane waves ($q\to\infty$): we call this light field a \textit{Bessel-lattice beam}~$\psi^\text{BL}_{\ell,q}$. Figs.~2b and c show the corresponding results for deltoid and cardioid caustics, respectively. We use a cw laser beam with wavelength $\lambda = 2\pi/k = \unit[532]{nm}$, shaped by a SLM, for their experimental realization. All beams demonstrated here have the same real space structure size $a=2\pi/k_\bot=\unit[15]{\upmu m}$ related to the propagation constant $k_z$ via $k_0^2 = k_\bot^2 + k_z^2$. 

Figure~2d presents the experimental verification of the propagation invariance of the astroid caustic, by showing a $xz$-cross-section of the measured intensity volume (corresponding to the white dashed line in Fig.~2 a5). 
This Bessel-lattice beam with charge $\ell = 0$ propagates for one Rayleigh length $z_e = 2ka^2 = \unit[5.32]{mm}$ without significantly changing its transverse intensity distribution.
In general, all propagation-invariant beams realizable with our method have standard properties with respect to their diffraction length: theoretically these beams do not diffract, but experimental limitations such as finite apertures lead to finite but very long focal lengths~\cite{Bouchal2003a, Durnin1987, Rose2012}. 

\begin{figure}[!t]
\centering
\includegraphics[width=1\columnwidth]{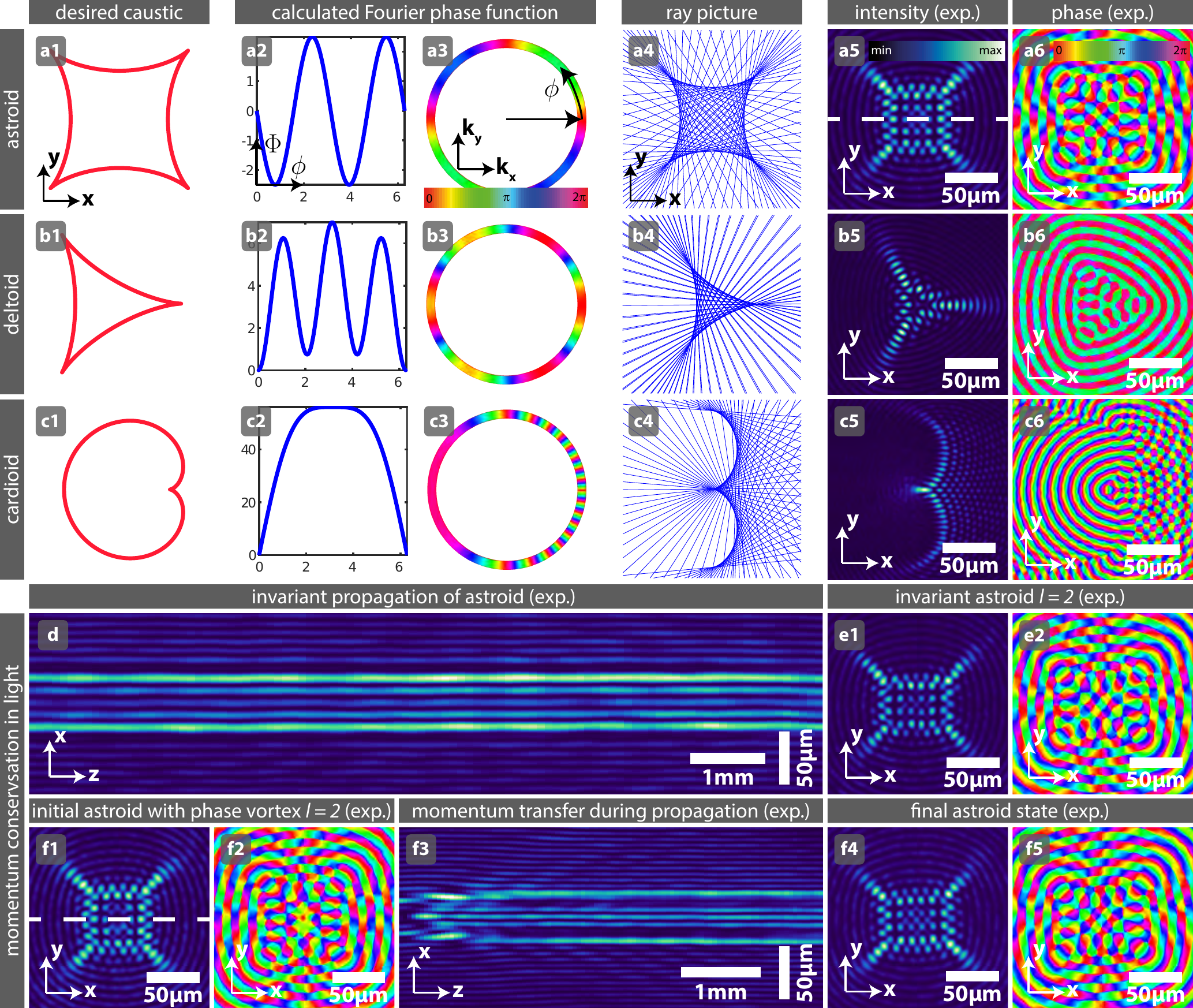}
\caption{Engineering desired caustics in light, demonstrating their propagation invariance and momentum conservation. 
   (a) Astroid. (a1) Desired caustic. (a2) 1D phase function. (a3) Phase function mapped onto 2D Fourier ring. (a4) Transverse projection of the rays. (a5, a6) Experimentally measured transverse intensity and phase. 
   (b) Deltoid. (c) Cardioid. (d) Invariant propagation of the astroid from image (a), $x$-$z$-cross section through intensity volume. (e) Invariant astroid with charge $\ell=2$. 
   (f) Self-healing / Momentum transfer in astroid: (f1), (f2) Diffractive astroid, initial field with phase vortex of charge $l=2$. (f3) Propagation showing diffraction and transfer to final OAM state. (f4) Momentum transfer to final astroid state same as (e), which then propagates invariantly.}
\label{fig:InverseCausticODE}
\end{figure}

Figure~2e shows a propagation-invariant Bessel-lattice beam $\psi^\text{BL}_{2,5}$, where the OAM with charge $\ell = 2$ has the effect of making the square lattice rectangular. 
An interesting 
effect
can be explored in which the Bessel-lattice beam $\psi^\text{BL}_{0,5}$ from~(a) is made to pass through a $\ell = 2$ phase vortex, resulting in the diffracting field $\psi = \psi^\text{BL}_{0,5} \cdot \exp\left[\text{i}\ell\phi\right]$. 
Its experimentally obtained initial intensity and phase distributions are shown in images (f1) and (f2). A cross-section of the intensity under propagation is shown in (f3), were we can see that the beam stabilizes during evolution over one Rayleigh length $z_e$ and acquires the shape corresponding to $\psi^\text{BL}_{2,5}$ as shown in (f4).
Hence, the total momentum of the light field is conserved and transforms the initial (diffractive) state to its invariant form when propagating sufficiently far~(g4, g5).
In the Supplementary Material we present further investigations and a generalization of this effect, which can be considered both as angular momentum transfer as well as a form self-healing~\cite{Bouchal2003a, Anguiano-Morales2007, Rubinstein-Dunlop2017} .

The first approach described above allows finding the phase functions that produce different caustic shapes. However, a limitation of this approach becomes evident from the example of a cardioid caustic in Fig.~2c. 
Since each point of the Fourier ring contributes to one and only one caustic point, the caustic is either convex or at least the curvature always must have the same sign. 
That is, inflection points of the desired caustic cannot be included in the intensity distribution, and even regions of small curvature give very faint intensity features~(c5). 
These limitations are overcome by the second approach described in what follows. More information on the algorithm and the experimental setup are given in the Methods.

\subsection*{Solving the inverse problem using Bessel beams as a pencil}

\begin{figure}[!t]
\centering
\includegraphics[width=1\columnwidth]{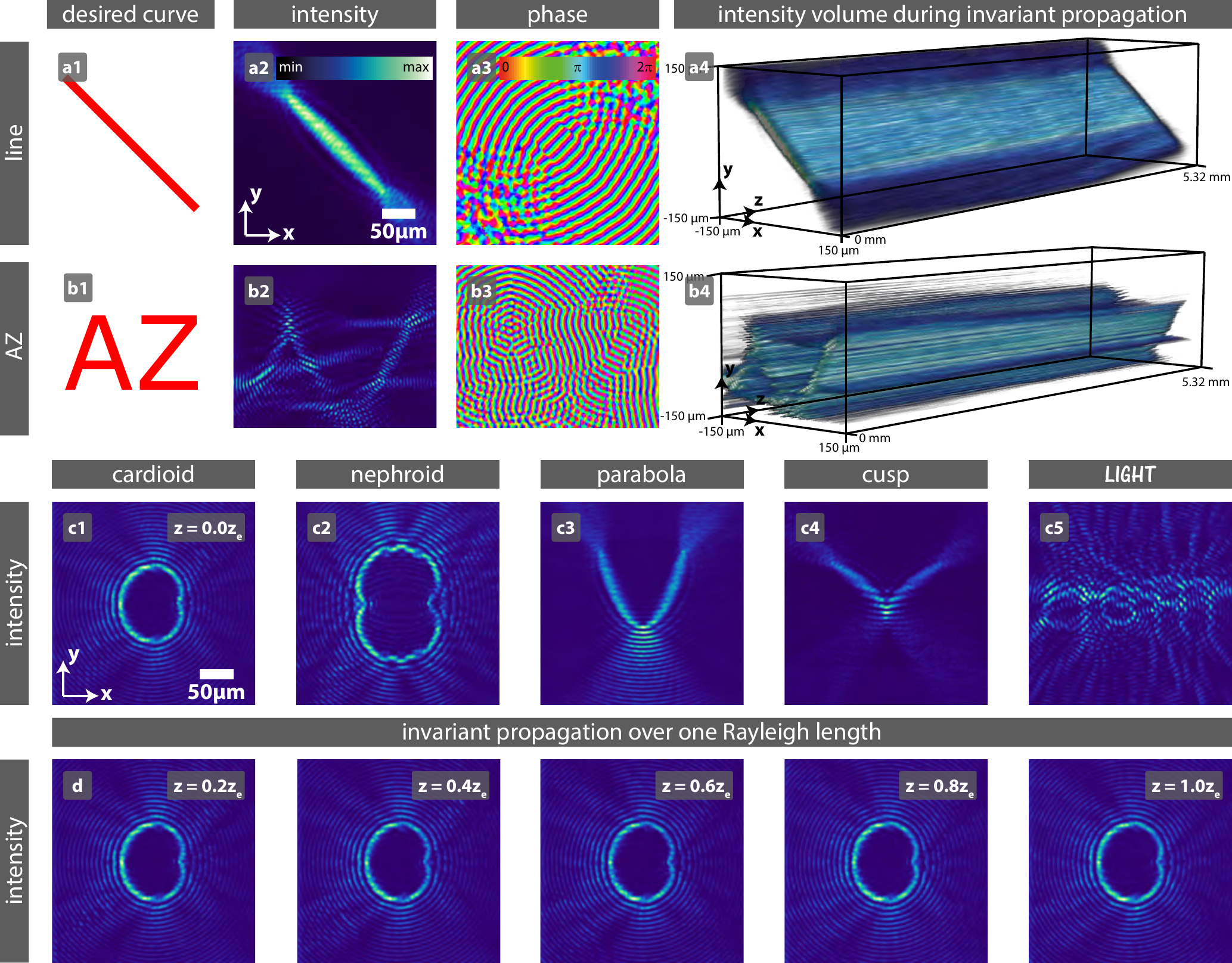}
\caption{Bessel pencil method for drawing any desired curve. 
   (a) Straight line segment. (a1) Desired curve. (a2), (a3) Experimentally measured transverse intensity and phase. (a4) Experimentally obtained intensity distribution over one Rayleigh length. 
   (b) Superposition of building blocks allows creating cuspoid letters 'AZ' in propagation-invariant light. (c) Smart beam design facilitates complex structures: Word 'LIGHT' in propagation-invariant light field. (d) Proof of invariance during propagation at the example of the cardioid (c1) over a distance of one Rayleigh length.}
\label{fig:BesselPencil}
\end{figure}

The second method for producing propagation-invariant intensity features in the transverse plane that follow any desired curve is based on \textit{drawing} this curve using a 0$^{\text{th}}$-order Bessel beam (whose caustic is a point) as a pencil.
The Methods section describes this algorithm.
Related approaches to create 3D high-intensity curves based on superimposing Gaussian beams or for lower-dimensional accelerating fields were demonstrated in~\cite{Rodrigo2013, Rodrigo2015, Wen2018}, but without the propagation-invariance achieved here.
Figure~3 shows a collection of experimentally obtained propagation-invariant beams generated with this algorithm.
A propagation-invariant straight line segment, measured over one Rayleigh \mbox{length $z_e$}, is shown in Fig.~3(a).
By deftly superimposing differently oriented cusps, the propagation-invariant letters 'AZ' can be composed over the transverse plane~(b).
The images~(c) show further propagation-invariant high-intensity rims in the initial transverse plane: a cardioid~(c1), a nephroid~(c2), a parabola~(c3), and a cusp~(c4) are examples of simple geometric shapes. Notice that the intensity is roughly uniform along the whole curve, regardless of the curvature. 
Smart beam design even allows for rather complicated shapes, as demonstrated as a proof of principle by imprinting the word `LIGHT' in the light field~(c5). The propagation-invariance of these and further beams created with the \textit{Bessel pencil method} is demonstrated at the example of the cardioid (c1) in the sequence of images in (d), which does not change its intensity over a propagation distance of one Rayleigh length. We show further examples of transversal curves and their invariance during propagation in the Supplementary Material.

\section*{Conclusion}

Caustics can be shaped into propagation-invariant light that presents any desired high-intensity curve in its transverse profile.
We present methods to tailor the light's wavefront, forming translation-invariant, customized caustics as the envelope of families of rays.
The two approaches presented here are illustrated by experimentally implementing propagation-invariant beams whose caustics trace fundamental forms like astroids, deltoids, cardioids, and nephroids, as well as more sophisticated structures such as letters or words.
These self-healing beams propagate robustly in the presence of perturbations. 
We demonstrate the properties of an astroid embedded as high-intensity caustic in form of a \textit{Bessel-lattice beam} that interpolates between plane waves and Bessel beams.
Propagation-invariant caustics satisfy the need for customized high-energy transfer in nano-fabrication applications with light or electron beams for ultrafast cutting and deep drilling in transparent materials~\cite{Mathis2012, Courvoisier2016}.
2D caustic light, e.g. with centred periodic lattices and well-defined curvilinear borders, enables fabricating refractive index modulations in (nonlinear) materials for novel topological structures.
Imaging systems like light sheet microscopy benefit from the robust long-focus propagation of high-intensity caustics with phase singularities~\cite{Vettenburg2014,Fahrbach2010}.

\section*{Methods}

\begin{figure}
\centering
\includegraphics[width=.75\columnwidth]{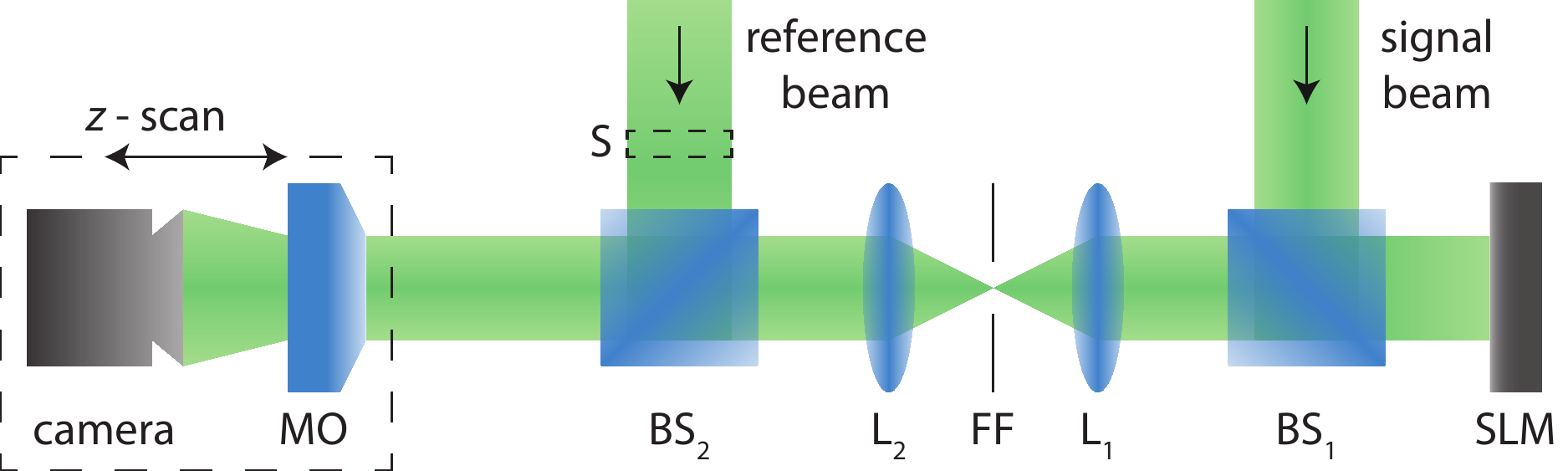}
\caption{Experimental setup. Nd:YVO$_4$ laser source, cw, wavelength $\lambda_0 = \unit[532]{nm}$. SLM: spatial light modulator. BS: beam splitter. L: lens. FF: Fourier filter. MO: microscope objective. S: shutter.}
\label{fig:Setup}
\end{figure}

\subsection*{Experimental setup}

Figure~\ref{fig:Setup} shows the experimental setup.
A frequency-doubled cw Nd:YVO laser beam with wavelength $\lambda_0 = \unit[532]{nm}$ is expanded in a plane wave, illuminating a phase-only full HD SLM `Holoeye HEO'. 
We address a pre-calculated phase pattern to the SLM that allows encoding the real space amplitude and phase simultaneously on a single phase-only modulator~\cite{Davis1999}. 
Therefore, an appropriate filter in Fourier space is necessary (FF). 
The desired light field establishes in the image plane of the modulator. 
A movable microscope objective (MO) and a camera constitute the imaging system, capable of observing the propagation of a light field. 
A second expanded beam, a plane wave, serves as a reference beam and can be switched on for phase measurements using a shutter (S). 
To recover the spatial phase, a standard digital holographic method is applied, that is based on the superposition of the signal beam with a slightly tilted reference beam~\cite{Schnars2005}. 


\subsection*{First approach} 
The caustics, parametrized by $\phi$, are located at the points $\mathbf{r}$ for which the first and second derivatives of the phase of the integral in Eq.~(\ref{eq:Whittaker}) with respect to $\phi$ vanish:
\begin{align}
   \Phi'(\phi) + k_\bot\mathbf{r}\cdot\mathbf{u}'(\phi) = 0, & & \Phi''(\phi) - k_\bot\mathbf{r}\cdot\mathbf{u}(\phi) = 0.
\end{align}
Since $\mathbf{u}$ and $\mathbf{u}'$ are orthonormal and complete over the plane, we can write $\mathbf{r} = (\mathbf{r}\cdot\mathbf{u})\mathbf{u} + (\mathbf{r}\cdot\mathbf{u}')\mathbf{u}'$ to find the parametrized caustic,
\begin{align}
   \mathbf{r}_c(\phi) &= \frac{1}{k_\bot}\left[\Phi(\phi)''\mathbf{u}(\phi) - \Phi'(\phi)\mathbf{u}'(\phi)\right].
\end{align}
Note that the angular coordinate $\phi$ in the Fourier ring is not an angular coordinate in the caustic space. 
A parametrization of the caustic with $\phi$ as the parameter may be found by considering the derivative of both sides of the previous equation,
\begin{equation}
   \mathbf{r}'_c(\phi) = \frac{1}{k_\bot}\left[\Phi'''(\phi) + \Phi'(\phi)\right]\mathbf{u}(\phi).
\end{equation}
Solving this differential equation allows for the realization of the light field $\psi(\mathbf{r})$ from Eq.~\eqref{eq:Whittaker} inversely, embedding in it the parametrized caustic $\mathbf{r}(\phi)$.

\subsection*{Second approach}

A second approach to realize a desired curve is to use the most localized propagation-invariant light spot we can achieve, a $0^\text{th}$-order Bessel beam, as a pencil to draw the curve. Since $\ell = 0$, its caustic is a point.
In accordance with Eq.~\eqref{eq:Whittaker}, the angular spectrum for such a Bessel beam centred at $\mathbf{r}_c$ is $A_\text{B}\exp\left[\text{i}\gamma_\text{B} - \text{i}k_\bot\mathbf{r}_c\cdot\mathbf{u}(\phi)\right]$. 
The key to design the real space light field $\psi(\mathbf{r})$ is to construct the angular spectrum by coherent integration of this expression along the desired path $\mathbf{r}_c(\theta)$ and choosing the phase $\gamma_\text{B}(\theta)$ and amplitude $A_\text{B}(\theta)$ appropriately along the curve.
Note that the parameter $\theta$ is not necessarily the angle $\phi$. 
Since the phase of the field along the curve should grow with the curve's arc length, it is given as
\begin{equation}
\gamma_\text{B}(\theta) = k_\bot\int_0^\theta \left|\mathbf{r}'_c(\tau)\right|\text{d}\tau.
\end{equation}
Similarly, to make the weight uniform along the curve we use
\begin{equation}
   A_\text{B}(\theta) \propto 1 / \sqrt{\left|\mathbf{r}'_c(\theta)\right|}.
\end{equation}
The total field can then be calculated via the angular spectrum as
\begin{equation}
   \psi(\mathbf{r}) = \int A_\text{B}(\theta) \exp\left[\text{i}\gamma_\text{B}(\theta) - \text{i}k_\bot\mathbf{r}_c(\theta)\cdot\mathbf{u}(\phi)\right] \text{d}\theta.
\end{equation}

{\bf Acknowledgements.}  
We are grateful for informative conversations with Danica Sugic and Eileen Otte.
Ollie Dyer performed some preliminary calculations with Bessel-lattice beams.

{\bf Funding.}  
M.A.A. acknowledges support from the National Science Foundation (PHY-1507278) and the Excellence Initiative of Aix-Marseille University- A*MIDEX, a French ``Investissements d'Avenir'' program.
M.R.D. gratefully acknowledges support from the Leverhulme Trust Research Programme RP2013-K-009, SPOCK: Scientific Properties of Complex Knots.

{\bf Competing interests.}  
The authors declare no competing interests.

{\bf Author contributions.}
C.D., M.A.A., and M.R.D. designed the project, developed the theory, and supervised the research. A.Z., C.D., and M.R.D. contributed to the experimental design and obtained the data. A.Z. drafted the manuscript. All authors analysed the data, defined the manuscript structure and worked on the manuscript.

\bibliography{Publications-10_Transverse_invariant_Caustics}

\cleardoublepage
\newpage
\setcounter{equation}{0}
\setcounter{page}{1}
\setcounter{figure}{0}
\renewcommand\thefigure{S\arabic{figure}}

{\fontsize{14}{14} \selectfont
\begin{center}
\textbf{Shaping caustics into propagation-invariant light \\| Supplemental Material}
\end{center}
}


%
%
%

\section*{Visualizing the rays of propagation invariant caustics}

We assume that the beams with propagation-invariant caustics propagate in linear, homogeneous, and isotropic media, so that their rays are straight lines. Caustics are defined as the envelopes of these rays~\cite{Berry1980, Blandford1986}. Each ray is described by its transverse coordinate \mbox{$\mathbf{Q} = (Q_x,Q_y)$} in the initial plane $z = 0$ and its transverse momentum $\mathbf{P} = (P_x,P_y)$. The momentum is given by the gradient of the 1D phase function $\Phi(\phi)$ mapped onto the Fourier ring in cylindrical coordinates. The evolution in $z$ of one ray is given by $\mathbf{Q} + z\mathbf{P}$~\cite{Alonso2017}.

For Bessel beams with topological charge $l \neq 0$ the envelope of rays describes a circular caustic, as shown in Fig.~1~(a) in the main part of this paper~\cite{Alonso2017}. The rays of the astroid beam form an envelope that is characterized by four cusps, shown in Fig.~1~\&~2 in the main manuscript. The left site of Fig.~S1 in this Supplemental Material shows the same subset of rays from several viewpoints, in order to image that the rays in the transverse plane emerge as mapping from three-dimensional rays to the plane. Recall that the full ray family is composed of a continuum of versions of the rays shown here displaced in the $z$-direction. This is demonstrated on the right site of Fig.~S1, where we add up, exemplary, 5 $z$-shifted subsets of rays (blue), forming an invariant caustic (red). Each iteration, the additional subset is highlighted by bold lines. The 5 discrete sets should give an impression of the continuous ray picture.   

\begin{figure}[]
\centering
\includegraphics[width=1\columnwidth]{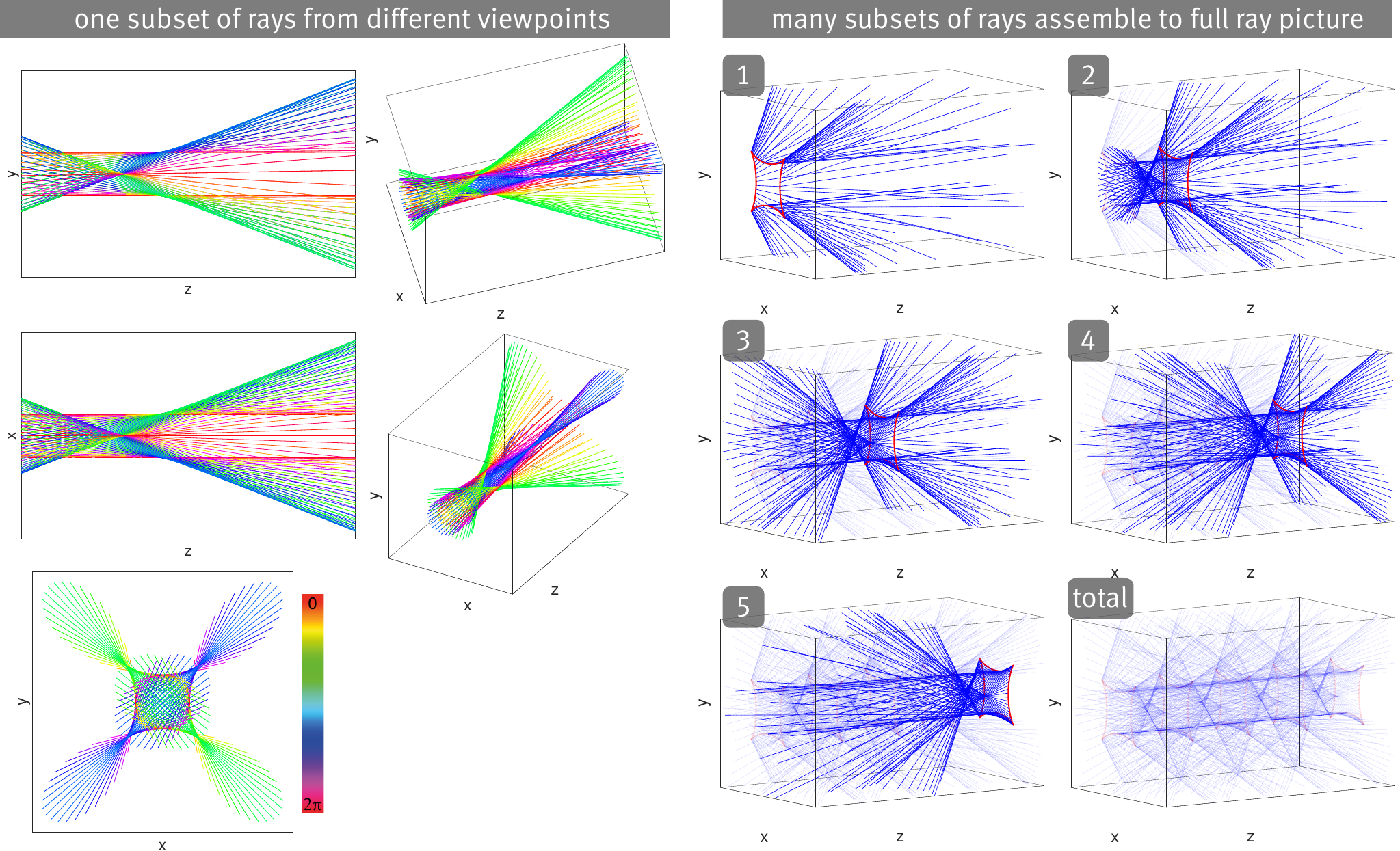}
\caption{Left: One subset of the ray picture of a tailored propagation-invariant astroid caustic. Color corresponds to phase. Shown are different angles of view to illustrate how this subset of the family of rays forms the caustic. Right: Assembly of the whole ray picture (total) by adding infinite many (exemplary shown are 5) z-shifted subsets of rays together. Rays are blue, caustics are red.}
\label{fig:Rays3D}
\end{figure}

\section*{
Generalized momentum transfer in propagation-invariant caustic beams}



We now discuss how a propagation-invariant beam can be made to morph into another by making it pass through a transparent mask with the appropriate azimuthally-dependent phase. 
Consider first the example of a propagation-invariant Bessel-lattice beam $\psi^\text{BL}_{\ell,q}(x,y)$ shown in Fig.~2  of the main text. As shown there, by modifying its phase with a vortex mask of charge $m = 2$, the field acquires  an intensity pattern that changes under propagation until it settles (for $z > z_e$) into the propagation-invariant field (at least in the center) of a corresponding Bessel-lattice beam with changed topological charge $\psi^\text{BL}_{\ell + m,q}(x,y)$. This process is represented symbolically as
\begin{equation}
\lim_{z\rightarrow\infty} \psi^\text{BL}_{\ell,q}(x,y,z) \cdot \exp\left[ \text{i}m\theta \right] \rightarrow \psi^\text{BL}_{\ell+m,q}(x,y),
\label{eq:BLMometumTransfer}
\end{equation}
where $\theta$ is the azimuthal angle. 

We can generalize this behaviour to any initial and final propagation-invariant beams, $\psi^i$ and $\psi^f$, whose phase functions are given by $\Phi^i(\theta)$ and $\Phi^f(\theta)$, respectively. 
%
The transition can be then implemented by using a transparent mask whose phase is $\Phi^f(\theta)-\Phi^i(\theta)$, namely
\begin{equation}
\lim_{z\rightarrow\infty} \psi^i
\cdot 
\exp\left[\text{i}\left(\Phi^f-\Phi^i\right)\right] \rightarrow \psi^f.
\label{eq:GeneralizedMomentumTransfer}
\end{equation}

We demonstrate this effect by transferring the three fundamental caustic shapes from Fig.~2, the astroid, the cardioid and the deltoid, into each other during propagation, shown in Fig.~S2. (a1), (b1), and (c1) show simulated initial transverse intensity distributions of the corresponding electric fields, whose individual phase is manipulated. During propagation, the transverse intensity distribution changes. After one Rayleigh length $z_e$ the final shape is apparent (a2), (b2), (c2) and fully developed after two Rayleigh lengths (a3), (b3), (c3). The experimental results (a4-a6), (b4-b6), (c4-c6) agree with the simulations. Only the initial intensity distributions show deviations from the simulations since the spatial light modulator is not capable to realize all of the (spatial) high-frequencies of the phase distribution that fluctuates strongly and interferes destructively. 

\begin{figure}
\centering
\includegraphics[width=1\textwidth]{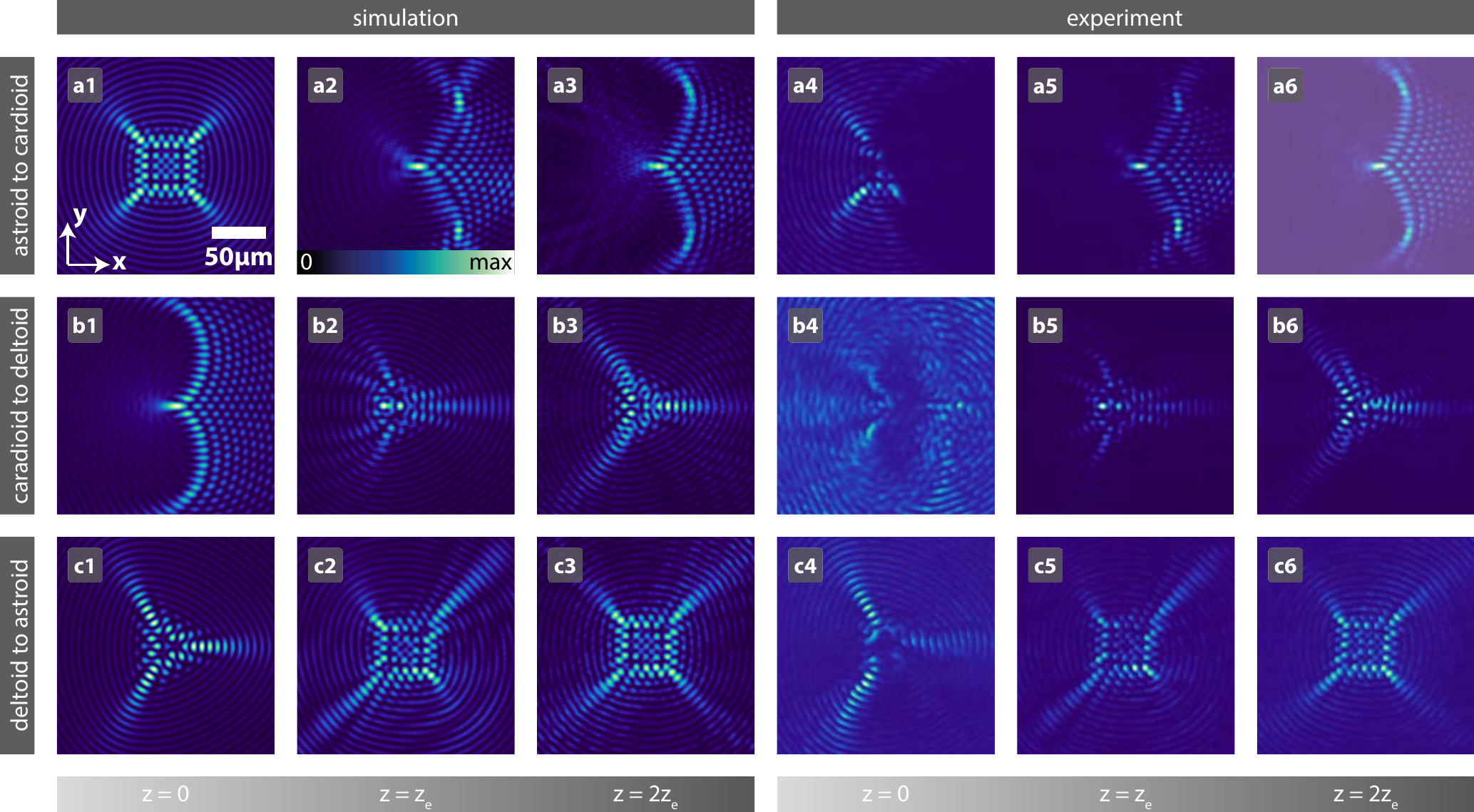}
\caption{Momentum transfers of an initial caustic to a final caustic by tailoring the initial phase. Compared are simulations with experiments. During propagation, (a) an astroid transforms to a cardioid caustic, (b) a cardioid transforms to a deltoid caustic, and (c) a deltoid transforms back to an astroid caustic. The propagation distances are 1 and 2 Rayleigh lengths $z_e$.}
\label{fig:Momentum}
\end{figure}

\section*{Bessel pencil for complex high-intensity features in \\propagation-invariant light}

\begin{figure}
\centering
\includegraphics[height=.85\textheight]{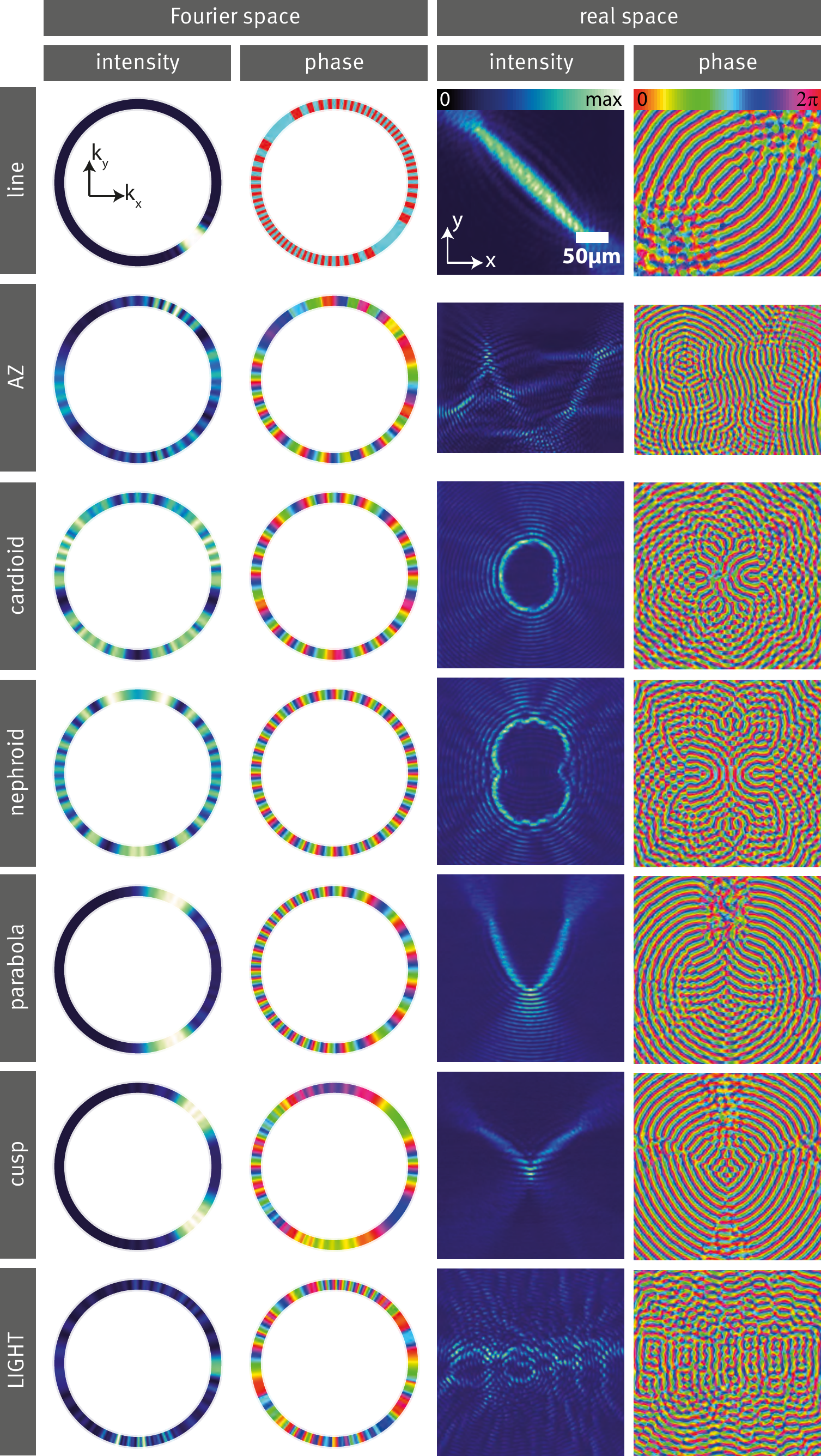}
\caption{Relation between the Fourier space intensity and phase distributions on a ring and the corresponding real space transverse intensities and phases for different curves realized with the \textit{Bessel pencil method}.}
\label{fig:BesselPencilFT}
\end{figure}

This section provides more experimental results and shows sophisticated high-intensity distributions realized with the \textit{Bessel pencil method}. We show the spectra of the beams in Fig.~S3, present several further examples whose high intensity curves describe both fundamental and rather complex shapes, and demonstrate their invariant propagation.

In order to create desired propagation-invariant high-intensity curves in the transverse plane, we integrate 0$^{\text{th}}$-order Bessel beams, whose caustics are points~\cite{Alonso2017}, coherently along these paths. We modulate both, the amplitude and phase of the spectrum, which is confined on a ring. Fig.~S3 shows the calculated intensity and phase distributions of the spectra and the corresponding propagation-invariant experimentally obtained transverse intensity and phase distributions.

All images in Fig.~S4 were obtained experimentally. The initial transverse intensity distributions can be seen in row 1, while row 2 shows the corresponding phases. The transverse intensity distributions after propagating one Rayleigh length $z_e = \unit[5.32]{mm}$ are shown in row 3. Row 4 shows 3D intensity volumes created from taking 50 transverse intensity distributions in equidistant steps from $z = \unit[0]{mm}$ to $z = z_e = \unit[5.32]{mm}$.

We demonstrate a deltoid (a), and astroid (b), which are hypocycloids with 3 or 4 cusps, respectively. They are followed by a hypocycloid with 5 cusps (c). From the class of epicycloids, we realized the cardioid (d) and the nephroid (e). 
Further, a parabola (f) and a cusp (g) may act as building blocks for rather complex shapes, such as words. 
For example, we create the letters LIGHT in a propagation invariant light field, demonstrating the potential of our approach.

\begin{figure}
\centering
\includegraphics[height=.8\textheight]{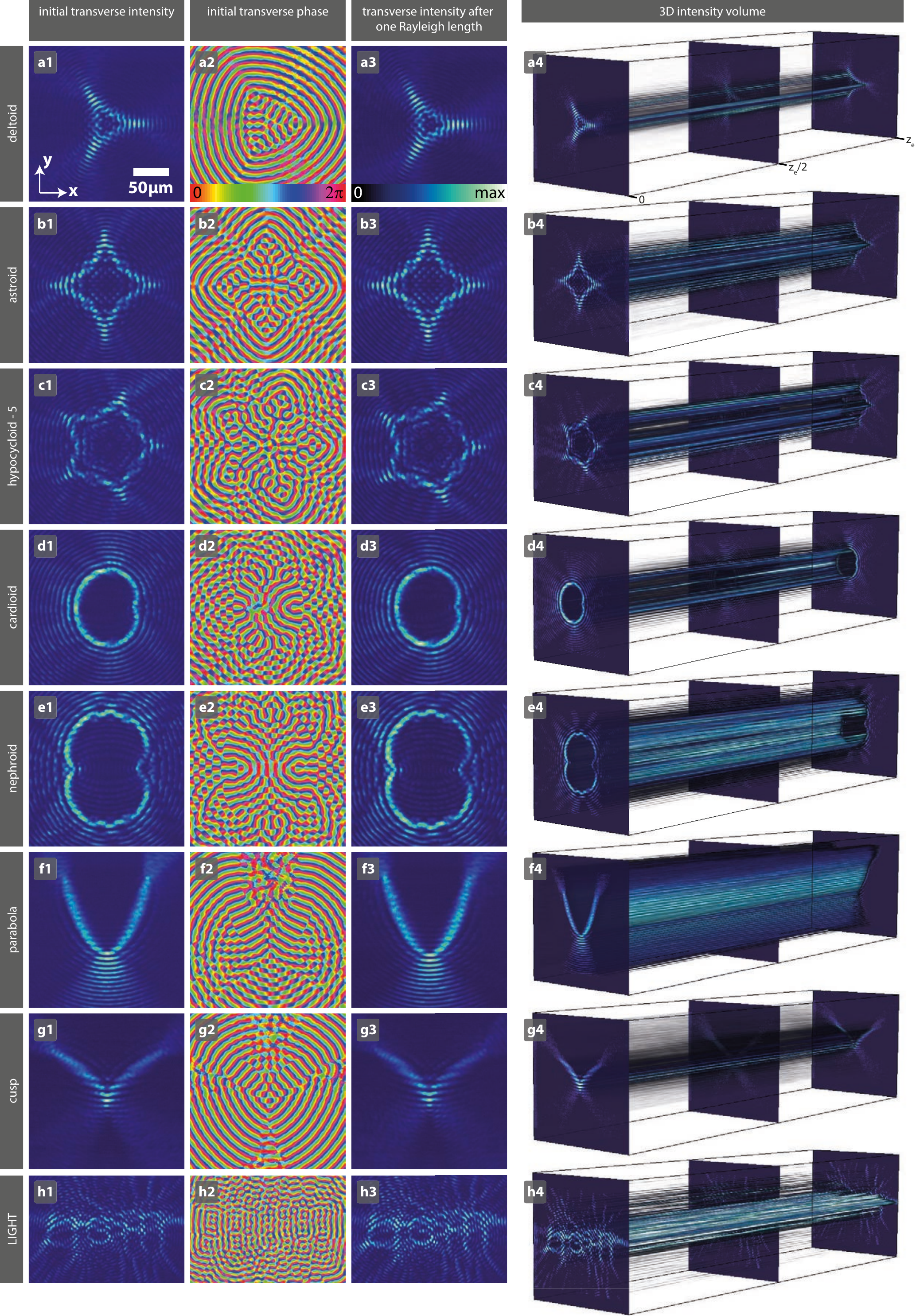}
\caption{A collection of experimentally-realized propagation-invariant beams realized with the \textit{Bessel pencil method}, whose profiles trace the following shapes: deltoid, astroid, hypocycloid with 5 cusps, cardioid, nephroid, parabola, cusp, and letters forming the word LIGHT. The initial transverse intensity and phase distributions are shown in rows 1 \& 2, respectively. Row 3 shows the (largely unchanged) transverse intensity after propagating one Rayleigh length. Row 4 shows the experimentally obtained 3D intensity volume.}
\label{fig:MoreLightFields}
\end{figure}

\end{document}